\documentclass[twocolumn,showpacs,amsmath,amssymb]{revtex4}
\usepackage{graphicx}
\usepackage{hyperref}
\usepackage{bm}
\begin{document}
\title{The Generation of Turbulence by Oscillating Structures in Superfluid Helium\\ at Very Low Temperatures}
\author{R.~H\"anninen}
\affiliation{Department of Physics, Osaka City University, Sumiyoshi-ku, Osaka 558-8585, Japan; and Low Temperature Laboratory, Helsinki University of Technology, PO Box 2200, FIN-02015 HUT, Finland.}
\author{M.~Tsubota}
\affiliation{Department of Physics, Osaka City University, Sumiyoshi-ku, Osaka 558-8585, Japan.}
\author{W.~F.~Vinen}
\affiliation{School of Physics and Astronomy, University of Birmingham, Birmingham B15 2TT, UK.}
\date{\today}
\begin{abstract}

The paper is concerned with the interpretation of many experiments that have been reported recently on the production of quantum turbulence by oscillating spheres, wires and grids in both $^4$He and $^3$He-B at temperatures so low that there is a negligible fraction of normal fluid. The experimental results are compared with those obtained in analogous experiments with classical fluids and with preliminary simulations of the quantum turbulence. Particular attention is paid to observed values of drag coefficients and to the very different critical velocities observed in $^4$He and $^3$He. It is tentatively concluded that in the case of $^4$He behaviour may well be similar to that observed in the classical analogues, with relatively small changes when the characteristic size of the oscillating structure is not large compared with the quantized vortex spacing, but that in the case of $^3$He behaviour is very different and due perhaps to very rapid intrinsic nucleation of the quantized vortices.    

\end{abstract}
\pacs{67.40.Vs, 47.27.Cn}
\maketitle
\section{Introduction}
Although turbulence in a superfluid $^4$He was discovered over fifty years ago, interest in it has recently intensified \cite{vinen1}, especially in connection with forms of superfluid turbulence in both $^4$He and $^3$He that have classical analogues. Flow of a superfluid is strongly influenced by quantum effects,  and the study of superfluid turbulence (or \textit{quantum turbulence}) is often concerned with the way in which these effects influence various forms of turbulence.  The quantum effects are of two types:  those leading to \textit{two-fluid behaviour}, a \textit{normal fluid}, behaving like a conventional viscous fluid, coexisting with an inviscid \textit{superfluid component}; and those that lead to quantum restrictions on the flow of the superfluid component. In pure $^4$He or pure $^3$He the normal component is composed of thermal excitations,  and it disappears at very low temperatures \cite{tilley1}. The behaviour of a  superfluid at very low temperatures is therefore of special interest, since it is concerned with the simple and fundamentally important case when there are no complications due to the presence of a normal component.  This paper will be concerned with this important case.  

Superfluidity is associated with the formation within the fluid system of a coherent particle field,  due to Bose condensation of helium atoms in the case of $^4$He and to BCS condensation of Cooper pairs of atoms in the case of $^3$He \cite{tilley1}. The long-range phase coherence of the particle field (or condensate wave function) leads to two restrictions on the velocity field, $\mathbf{v}_{s}$, of the flowing superfluid component: in a simply-connected volume  curl$\bf{v}_{s}=0$; and in a multiply-connected volume the circulation is subject to quantization in the form
\begin{equation}  \oint \mathbf{v}_{s}\cdot dr = n\kappa,
\label{eq1}
\end{equation}  
where $n$ is an integer, $\kappa=2\pi\hbar/m$ is the quantum of circulation, and $m$ is the mass of a single $^4$He atom or a pair of $^3$He atoms.  Free vortex lines can exist in the bulk of the superfluid,  provided that the line has a core at the centre of which the condensate wave function vanishes. In practice free vortex lines in both $^4$He and the low-temperature phase of $^3$He, $^3$He-B, have single quanta of circulation. The radius of a core is typically about 0.05 nm in $^4$He and 80 nm in $^3$He-B.  

Since turbulent flow is necessarily rotational, turbulence in the superfluid component can arise only through the presence of quantized  vortex lines, and it must take the form of what is often described as a random tangle of lines. This term is rather misleading because in practice the tangle is often not random,  but rather is locally polarized in such a way as the produce velocity fields on scales much larger than the spacing between the vortex lines. Indeed this polarization is often crucially important in forms of turbulence with classical analogues, because it allows flow on a wide range of length scales, as is characteristic of many forms of classical turbulence \cite{vinen3}.    

The study of classical turbulence has often been concerned with flows that are on average steady, either through a grid, or past an obstacle such as a cylinder or a sphere. Grid flow produces a particularly simple form of turbulence, since, well behind the grid, it is approximately homogeneous and isotropic \cite{batchelor1}. Steady flow past an obstacle relates to many practical problems. Controlled steady flow of a superfluid is difficult to achieve,  although turbulence in the wake of a steadily moving grid in $^4$He at relatively high temperatures has been the subject of very important studies \cite{stalp1}. Steady flow at very low temperatures is especially difficult to achieve, although plans to study turbulence in the wake of a moving grid are well advanced. It is,  however, rather easy to produce oscillatory motion of an obstacle or a grid in a superfluid at a very low temperature, and it is with the results of such experiments that this paper is concerned.  Unfortunately flow associated with such oscillatory motion is quite complicated, even in the classical case, and this makes for difficulty in interpreting experiments in the quantum case. Nevertheless, useful progress in this interpretation has already been made, and our paper aims to extend this progress. Sometimes we shall argue that effects have close classical analogues;  in other cases the quantum case seems rather different.

As is well-known, the creation of vortex line in a flowing superfluid is inhibited by a potential barrier, analogous to that opposing condensation of a supersaturated vapour \cite{vinen2}. In the case of $^4$He this barrier is so high that it cannot be overcome at low temperatures,  by either tunnelling or thermally,  unless the flow velocity is greater than $\sim$10 ms$^{-1}$. In all the cases that we shall examine the velocities are much too small, so that nucleation of turbulence must be extrinsic, relying on the growth of remanent vortices that are left over from previous turbulent flows or formed during cooling through the superfluid phase transition. In the case of $^3$He-B the barrier is much smaller, owing to the much larger core size, and intrinsic nucleation may be relevant to the experiments that we shall examine. 

It has become clear during the past decade that quantum turbulence  can often mimic its classical counterpart, the most striking example being the Richardson cascades and Kolmogorov energy spectra \cite{frisch1} observed in appropriate quantum cases, especially those involving homogeneous turbulence \cite{stalp1,vinen1,vinen3}. We shall examine the extent to which experimental evidence points also to similarities in flow past obstacles at very low temperatures. We shall find that such similarities do seem to exist, at least in part, with $^4$He,  but not apparently with $^3$He-B, and we shall try to discuss why this difference exists. We shall report the results of preliminary simulations of quantum turbulence that are relevant to our discussion; we shall show that these simulations already provide interesting hints, but that further progress will depend on the development in future of simulations that are more extensive and more time-consuming. This development will take time, and we take the view that publication of our results so far should not await its completion.

Our paper is arranged as follows. In Section II we shall summarize the experimental results that have been reported on the behaviour of oscillating spheres, wires and grids in both $^4$He and $^3$He-B, and we shall emphasize both certain common patterns of behaviour and certain differences between the behaviour of the two isotopes. In Section III we shall summarize behaviour observed with classical fluids. In Section IV we shall compare the classical and quantum cases for $^4$He, basing our knowledge of the quantum case on both experiment and the results of the preliminary simulations. Section V will be concerned with a parallel discussion for $^3$He-B, and Section VI will be devoted to summaries of our present understanding and of further work that is required.

\section{Experimental Results on Oscillating Structures in $^4$He and $^3$He-B at very low temperatures}

\subsection{Critical velocities}

\begin{figure}[ht]
\begin{center}
\includegraphics[width=0.4\textwidth]{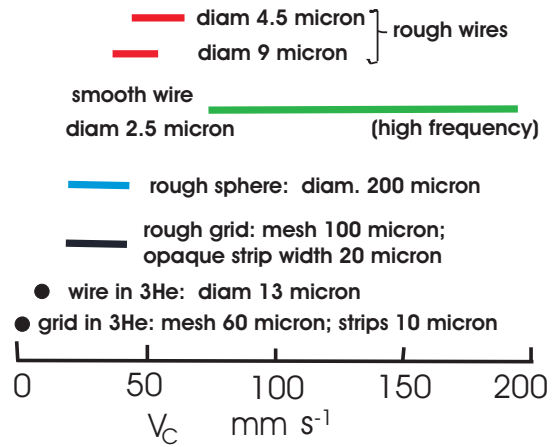}
\caption{Critical velocities for various oscillating structures. The bars relate to $^4$He. }
\label{fig1}
\end{center}
\end{figure}

Many experiments have now been reported, and we aim to summarize the principal findings. In all cases it is observed that the drag on the oscillating structure is consistent with ideal potential flow of the superfluid component at low velocity amplitudes, but that above a critical velocity there is increased drag, the magnitude of which tends at high velocities to be proportional to the square of this velocity.  Fig.~\ref{fig1} summarizes the observed critical velocities. Observations on a 100 $\mu$m radius sphere in $^4$He  were reported by Schoepe's group \cite{schoepe1,schoepe2,schoepe3,schoepe4}; those on vibrating wires in $^4$He by Bradley \textit{et al} \cite{bradley1} and by Yano \textit{et al} \cite{yano1}; those on vibrating grids in $^4$He by Nichol \textit{et al} \cite{nichol1,nichol2} and by Charalambous \textit{et al} \cite{charalambous1}; those on vibrating wires in $^3$He-B by Fisher \textit{et al} \cite{fisher1}; and those on a vibrating grid in $^3$He-B by Bradley \textit{et al} \cite{bradley2,bradley3}.         

The precise critical velocity varies a little from one experiment to another, as indicated by the bars in Fig.~\ref{fig1}, and especially for wires there is significant hysteresis (discussed in more detail below), the extent of which contributes to the length of the bars (there is very large hysteresis for the thin smooth wire). In the case of wires there is evidence that the critical velocity increases with increasing frequency, especially above about 2 kHz, and the relatively large values seen with a smooth wire are associated with the fact that they relate to a high frequency (about 3.8 kHz,  compared with about 1 kHz or less in the other cases). We see that for $^4$He the values lie within a surprisingly small range,  which seems independent of the type of structure and of the character of the roughness of the surface of this structure (the roughness of the grids is very different in character from that of the wires). Since, as we have noted, the critical velocities in $^4$He must be extrinsic, depending on remanent vortices, it is surprising that changes in roughness appear to have such a small effect; we would expect increased roughness to increase the density of remanent vortices and hence decrease the critical velocity.  Very recently Yano (private communication) has suggested that the critical velocity in $^4$He should be defined, in terms of a plot of the velocity against drag force, as the velocity at which the line describing the drag in the potential-flow state (due to extraneous damping and residual phonons) intersects the line describing the drag in the fully turbulent state, this latter line being if necessary extrapolated backwards (we discuss in an Appendix the assumptions that underlie definition, which are not necessarily justified). With this definition Yano shows, remarkably, that all the critical velocities, plotted as a function of frequency, lie on a single curve, with the critical velocity being roughly proportional to the one third power of the frequency at the lowest temperatures.      

Critical velocities in $^3$He-B are noticeably smaller than those in $^4$He, especially with a vibrating grid, for which the critical velocity is not well-defined but is not greater than about 1 mm s$^{-1}$. Furthermore, they may be intrinsic. Those observed with a vibrating wire are similar to those at which pair breaking occurs (about 8 mm s$^{-1}$), and pair breaking and vortex creation seem to occur simultaneously. In the case of a vibrating grid there is evidence that pair breaking is unimportant in the relevant range of velocities, but, as we shall argue later, it is at least possible that vortex creation is indeed intrinsic and does not rely on remanent vortices.   

In the cases of an oscillating sphere or a vibrating wire,  the transition to turbulence is accompanied by what can be described loosely as hysteresis. However, in discussing these effects we must remember that the velocity response is measured as a function of drive amplitude, so that for each experimental point a constant amplitude of drive is imposed. This is in contrast to the simulations that we report later, in which the constant velocity amplitude is imposed.  With a constant drive there can be a switching phenomenon, in which the velocity amplitude builds up until it exceeds the critical value for transition to turbulence, whereupon the increased dissipation leads to a reduction in the velocity amplitude to a value below critical; the decreased dissipation then leads to a rise in the velocity amplitude above critical, and so on. It appears that this switching occurs most straightforwardly over a range of drives in the case of the oscillating sphere studied by Schoepe \textit{et al}. This means that if each experimental point were obtained at constant velocity amplitude there would be no hysteresis. In the case of a vibrating wire switching seems sometimes to occur, but not always; when it does not occur straightforward hysteresis is observed, an increasing drive amplitude leading to a transition to turbulence at a higher velocity that that obtaining with a decreasing drive amplitude. It is possible of course that switching requires more time in the case of the vibrating wire,  and that it would therefore always be observed if the experiments were conducted more slowly. There is little evidence for either switching or hysteresis in the most recent (and probably most reliable) work on oscillating grids \cite{charalambous1,bradley2,bradley3}.         

\subsection{Drag Coefficients}

The proportionality of the supercritical drag to the square of the velocity (at least at high velocities) has led authors to a comparison with the drag observed in classical fluids at high Reynolds number \cite{batchelor2}. For steady flow at velocity $U$ past an obstacle in a classical fluid of density $\rho$ the drag can be written in the form

\begin{equation} F=\frac{1}{2}\rho U^2 C_D A,
\label{eq2}
\end{equation}
where $A$ is the projected area of the obstacle normal to the flow, and $C_D$ is a dimensionless \textit{drag coefficient}. We shall discuss values of classical drag coefficients in Section III; here we note that for a classical fluid at high Reynolds number this value is of order or a little less than unity. These classical drag coefficients relate to steady flow past an obstacle, but, as we shall see in Section III, they are relevant also to oscillating flows.

We can estimate drag coefficients for the quantum cases, in the limit of high velocities, from the published data on the velocity-dependent drag (Table I).  For the case of a flow that oscillates in time (or  of an obstacle that undergoes oscillatory motion in an otherwise stationary fluid) we need to take care in defining $C_D$: again we base it on Eq.~(\ref{eq2}), but with $U$ and $F$ taken as peak values. In cases where not all parts of the structure oscillate with the same amplitude we assume that Eq.~(\ref{eq2}) holds for each element of the structure, and then we average over the structure. For the stretched grid in $^4$He studied by Charalambous \textit{et al} \cite{charalambous1} we have carried out this averaging carefully, taking into account the fact that the spatial dependence of the displacement is a zero-order Bessel function; in the case of wires, lack of information has led us to make only rough estimates,  which are good to only perhaps a factor of two. We find that for $^4$He the drag coefficient is always of order or a little less than unity for the fully-developed turbulent regime,  but that for $^3$He-B it seems to be always much larger than unity;  in the case of a grid it is larger by a factor of at least 20. In the case of a wire in $^3$He-B  part of the drag may be due to the pair breaking that seems to accompany the turbulence, so that the figure of 9.5 in Table I is unreliable. We shall attach importance in our later discussion to the different values of the drag coefficient in the two isotopes. 

\vspace{0.5cm}
\begin{center}
TABLE I

\vspace{0.5cm}
\begin{tabular}{|l|l|l|}
\hline
Superfluid & Oscillating structure & $C_D$\\
\hline
\hline
$^4$He & Sphere, radius 100$\mu$m & 0.36\\
$^4$He & Wire, radius 1.25$\mu$m & 0.17\\
$^4$He & Wire, radius 2.25$\mu$m & 0.13\\
$^4$He & Grid, strip width 20.8$\mu$m & 0.29\\
\hline
$^3$He-B & Wire, radius 6.5$\mu$m & 9.5\\
$^3$He-B & Grid, strip width 11$\mu$m & 17.8\\
\hline
\end{tabular} 
\end{center}

\subsection{Observations of grid flow in $^3$He-B}

In the case of $^4$He no technique has yet been developed to study the form of the superfluid turbulence produced in the wake of a moving obstacle, except for the application of Particle Image Velocimetry to very large obstacles in a thermal counterflow at high temperatures, where the interpretation is made difficult by the presence of the two fluids \cite{vansciver1}. However, in the case of $^3$He-B at low temperatures the Lancaster group has been developing a technique that promises to provide us with pictures of superfluid turbulent fields, and which is based on the Andreev scattering of thermal quasi-particles from a superfluid velocity field \cite{fisher1,bradley2}. The interpretation of the observations is not quite straightforward, and we shall not discuss it here. Nevertheless, the following tentative picture has emerged of the form of turbulence produced in the wake of an oscillating grid in $^3$He-B.  At least at low velocities vortex rings emerge from the grid, presumably in all directions, and at sufficiently small velocities these rings interact to a negligible extent and simply fly away from the grid. At higher velocities the density of rings has increased to the point where interactions become important, and this leads apparently to a turbulent tangle of lines which remains localized in the neighbourhood of the grid. We shall comment on this picture later in our discussion.   

\section{Oscillating structures in a classical fluid}

In this section we aim to summarize what is known about the form of turbulence produced by oscillating spheres, cylinders and grids in a classical fluid. We are interested in behaviour at high Reynolds number, and we shall start with the case of cylinders, about which most seems to be known. 

\subsection{Cylinders}
\begin{figure}[ht]
\begin{center}
\includegraphics[width=0.4\textwidth]{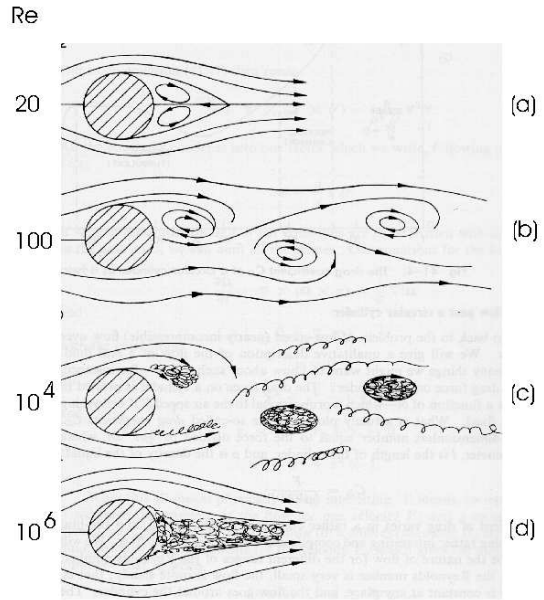}
\caption{Patterns of flow in the wake of a cylinder moving with constant velocity in a classical fluid \cite{feynman1}.}
\label{fig2}
\end{center}
\end{figure}

We recall first the behaviour of a cylinder of diameter $d$ moving in a classical fluid with \textit{constant velocity}. There is then one characteristic dimensionless number, namely the Reynolds Number, Re$=dU/\nu$, where $\nu$ is the kinematic viscosity of the fluid \cite{batchelor2}. For very small values of Re there is laminar flow, and the drag on the cylinder is given by a formula due to Lamb \cite{batchelor2} (the analogue of Stokes law for a sphere). If the flow is set up at high Reynolds number, the vorticity is initially confined to a thin \textit{boundary layer}, flow outside the boundary layer being irrotational. The (Bernouilli) pressure distribution associated with this potential flow is such that initially it leads to no net force on the cylinder (the d'Alembert paradox). The pressure is a maximum on the stagnation lines at the front and rear of the cylinder, and it falls to a minimum value in between. The flowing fluid at the rear of the cylinder therefore experiences a rising pressure as it moves towards the rear stagation line. This causes the fluid within the boundary layer to move backwards and give rise to the phenomenon of separation, so that a region is formed behind the cylinder where the flow is rotational. The resulting flow at a Reynolds number of order 20 is shown in the upper diagram in Fig.~\ref{fig2} \cite{feynman1}. As the Reynolds number increases the pattern of rotational flow evolves, as shown in the lower diagrams of Fig.~\ref{fig2}. When Re$>\sim 100$, vortices start to be shed from behind the cylinder,  and a gradually increasing Reynolds number leads eventually to a wake in which there is fully developed turbulence.  
\begin{figure}[ht]
\begin{center}
\includegraphics[width=0.4\textwidth]{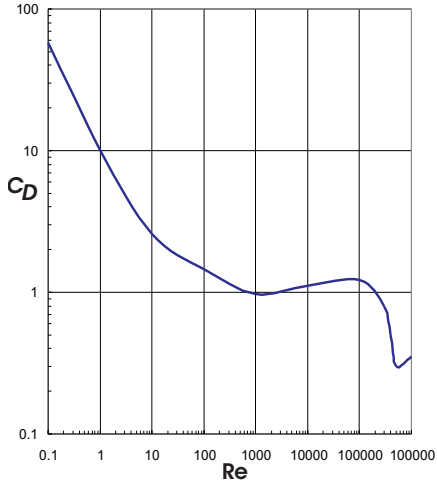}
\caption{Drag coefficient plotted against Reynolds number for flow past a cylinder}
\label{fig3}
\end{center}
\end{figure}

If we describe the drag on the cylinder as a function of Re by Eq.~(\ref{eq1}), we find that the drag coefficient $C_D$ varies as shown in Fig.~\ref{fig3} \cite{prandtl1}.  At large values of the Reynolds number the drag on the cylinder arises mostly from the pressure distribution at its surface. Within the separated wake the pressure is roughly constant and equal to that at the line of separation, the fluid velocity within the wake, near the surface of the cylinder, being significantly smaller than that in the potential flow. It follows that the pressure distribution gives rise to a force on the cylinder of order $(1/2)\rho U^{2}$ per unit area  projected on a plane normal to the motion of the cylinder \cite{batchelor2}. Therefore the drag coefficient $C_D$ is of order unity, as observed, although its precise value depends on the details of the flow and the position of the line of separation. We see from Fig.~\ref{fig3}, however, that there is no abrupt change in the behaviour of $C_D$ in the transition from laminar flow to turbulent flow. This suggests that the vortices depicted behind the cylinder at the top of Fig.~\ref{fig2} develop gradually, at least to some extent. For Reynolds numbers in the range $10^2$ to $10^5$ $C_D$ is very close to unity, but it drops to a value of about 0.3 at the so-called \textit{drag crisis}, when the turbulent wake suddenly becomes narrower because the boundary layer has become turbulent.

For an oscillating cylinder (amplitude of oscillation $a$; frequency $f$) the situation is more complicated because the single dimensionless number, Re, must be replaced by two numbers. The choice is to some extent arbitrary, but a commonly used pair are the Keulegan-Carpenter number, given by

\begin{equation} K_C=\frac{2\pi a}{d},
\label{eq3}
\end{equation}  
and the Stokes number, given by
\begin{equation} \beta=\frac{fd^2}{\nu}.
\label{eq4}
\end{equation}
We note that the Reynolds number is simply the product of $K_C$ and $\beta$.   

Patterns of turbulence generated by an oscillating cylinder have been observed by many authors. For our purposes we refer to the work of Williamson \cite{williamson1}, Obasaju \textit{et al} \cite{obasaju1}  and Sarpkaya \cite{sarpkaya1}, which relate to values of $K_C$ roughly in the range from 0.4 to 100 and to values of $\beta$ roughly in the range 100 to $10^4$. The observed patterns of flow are quite complicated and vary with the values of $K_C$ and $\beta$, and it is neither practicable nor appropriate at this stage in our work to describe these patterns in detail. Typically, however, motion of the cylinder during one half-cycle tends to produce a vortex pair of the type shown in the upper diagram of Fig.~\ref{fig2}, although the two vortices may be of unequal strength. At small values of $K_C$ these vortices do not become completely detached. They may be carried to the other side of the cylinder during flow reversal, but eventually they seem to be dissipated. At larger value of $K_C$ vortices do become detached, in pairs that propagate away from the oscillating cylinder; the pairing may involve vortices formed on opposite sides of the cylinder during successive half-cycles, and these pairs tend to propagate in a direction perpendicular to the direction of oscillation of the cylinder, often as a vortex street.  In view of the general tendency to produce vortex pairs behind the moving cylinder during one half-cycle we can expect the drag on the cylinder to have the form of Eq.~(\ref{eq2}), with $C_{D} \sim 1$.  This expectation is confirmed by the measurements reported in the papers by Sarpkaya and Obasaju \textit{et al}, who find that, at parameters large enough to cause the breakdown of laminar flow, $C_D$ lies always in the range between roughly 0.5 and 2, only slightly larger in general than is the case for steady flow. However, it should be emphasized that the drag coefficient is observed to vary with velocity within this range in a characteristic and roughly oscillatory way  that reflects changes in the detailed form of the flow.    

\subsection{Spheres}
As far as we know, there has been no published study of the flow associated with an oscillating sphere at high Reynolds numbers, comparable with those for cylinders. However, an experimental study is now in progress by Donnelly \cite{donnelly1}, and we can summarize the preliminary results obtained so far. Vortices are produced if $K_C> \sim 3$. The vortex pair that tends to be produced in the wake of the moving cylinder during one half-cycle is replaced by a vortex ring. This ring may move to the opposite side of the sphere during the next half cycle (performing a "leapfrog" motion over the sphere), but it is eventually shed by the sphere, so that a sequence of rings propagates away from the sphere in both directions along the line of motion. No regime in which vortices are not shed has yet been observed. Preliminary measurements of the drag lead to drag coefficients of order unity.

\subsection{Grids}

\begin{figure}[ht]
\begin{center}
\includegraphics[width=0.4\textwidth]{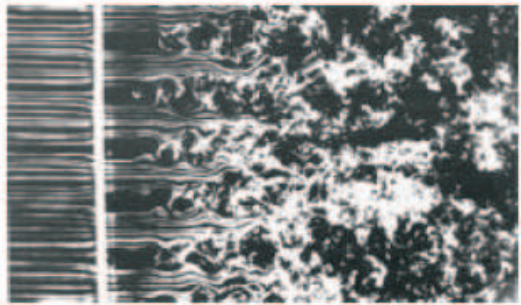}
\caption{Steady flow of a classical fluid through a grid \cite{frisch1}.}
\label{fig4}
\end{center}
\end{figure}

We consider a grid in the form of a thin solid sheet in which a regular pattern of holes has been drilled.  The pattern of flow through such a grid when it moves at a steady velocity perpendicular to its own plane is well-known and shown in Fig.~\ref{fig4}. A jet is formed at each aperture, the boundaries of the jet being vortex sheets. After a distance of order the mesh length the instability of these vortex sheets leads to the break up of the jets into vortex rings, which then interact to produce turbulence that is approximately homogeneous and isotropic. The drag on the grid must have the form of Eq.~(\ref{eq2}), with $C_D=1$,  where the area is that of the opaque portion of the grid. In some ways the opaque portions of the grid seem to behave like a disc placed normal to the flow,  for which the line of separation is at the edge of the disc,  and for which the observed drag coefficient is constant and very close in value to unity.  

A recent study of the flow pattern produced by a form of oscillating grid has been reported by Voropayev and Fernando \cite{voropayev1}. The grid consists of an planar array of circular cylinders, with mesh size $M$. Oscillation of this grid with velocity amplitude $U$ produces vortex pairs at each of the cylinders, each pair being similar to those produced by a single oscillating cylinder. These vortex pairs interact to produce a turbulent front, which diffuses away from the grid. The authors give a detailed analysis of this diffusion process, but the essential features seem to be as follows. If we assume that the turbulence spreads by diffusion, then the appropriate diffusion coefficient is probably given by something like
\begin{equation} D \sim \alpha U M = \alpha \nu \textrm{Re},
\label{eq5}
\end{equation}
where $\alpha$ is a constant less than unity that describes the factor by which velocity in the eddy motion is less than $U$, and Re is the Reynolds number $MU/\nu$. The turbulent front will therefore diffuse a distance $x$ in time $t$,  where
\begin{equation}  x^{2}=   (\alpha\nu \textrm{Re}) t.
\label{eq6}
\end{equation}
This result is closely similar to that derived by Voropayev and Fernando by more sophisticated methods. Since each cylinder composing the grid produces a vortex pair, we can expect that the drag will be given by Eq.~(\ref{eq2}), again with $C_D\sim 1$. Similar principles are likely to apply to other forms of grid, including one formed from holes drilled in a sheet.

\section{Oscillating structures in $^4$He at a very low temperature: discussion and simulations}

\subsection{Drag coefficients}
We have noted that the forces on oscillating structures in $^4$He are described at high velocities by Eq.~(\ref{eq2}), with a value of the drag coefficient that is a little less than unity.  This suggests that the behaviour of $^4$He may be similar, at least in some respects, to that of a classical fluid. We shall examine this idea in what follows, calling on the results of preliminary simulations that we describe in Section IVD.
\begin{figure}[ht]
\begin{center}
\includegraphics[width=0.4\textwidth]{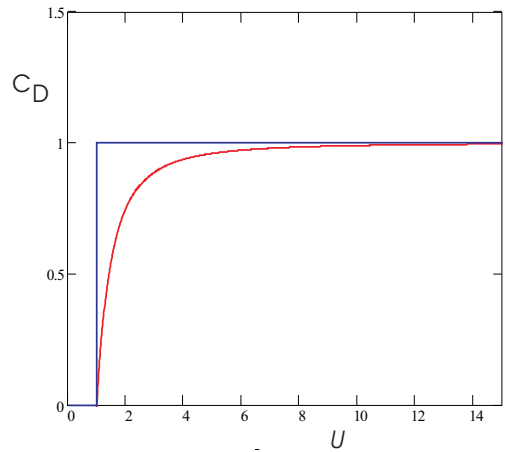}
\caption{Possible forms of dependence of the drag coefficient on velocity in $^4$He at zero temperature.}
\label{fig5}
\end{center}
\end{figure}

We note at this point that  in one important respect the behaviour of the drag coefficient for $^4$He must differ from that for a classical fluid.  This difference arises because at low velocities (analogous to low Reynolds numbers) the drag is zero for the superfluid (at zero temperature). This means that the velocity dependence of $C_D$ in $^4$He must differ from that in Fig.~\ref{fig3}. Two possible forms are shown in Fig.~\ref{fig5}. The studies of Schoepe \cite{schoepe3} show that for a sphere at zero temperature the dependence of drag force on velocity has the form

\begin{equation} F=\frac{1}{2}\rho C A U^2 - F_0,
\label{eq7}
\end{equation}
where $C$ and $F_0$ are constants, which is in fact the form shown in the lower, smooth, curve in Fig.~\ref{fig5}. For such a form to arise the rotational flow in the wake of the sphere must develop only gradually as the velocity of flow is increased through the critical value, as probably happens in the classical case. 

Values of the Keulegan-Carpenter number for the experiments on oscillating structures in $^4$He are as follows: for the sphere, $K_C \sim 1$; for the wires, $K_C \sim 10$; for the grid, $K_C \sim 2.5$; for $^3$He the values for wires and grids are smaller by factors of order 20.  

It is interesting to ask whether the experimental measurements on the various oscillating structures in a superfluid reveal a drag coefficient  at supercritical velocities that oscillates with velocity  by factors of order two, as is the case with a cylinder in a classical fluid, oscillations that are associated, as we have already noted, with changes in the detailed structure of the flow. As far as we have been able to judge, such oscillations are not present in the superfluid case; no such oscillations seem to be superimposed on a smooth variation of the type shown in Fig.~\ref{fig5} (see, for example,  the very careful measurements of Niemetz and Schoepe \cite{schoepe3}). If this judgement is correct, then there is evidence that, in contrast to the classical case, the pattern of turbulent superflow remains the same over a wide range of velocity. The similarity between the quantum and classical cases may therefore be limited.  

\subsection{The extrinsic character of the critical velocity in $^4$He}

\begin{figure}[ht]
\begin{center}
\includegraphics[width=0.3\textwidth]{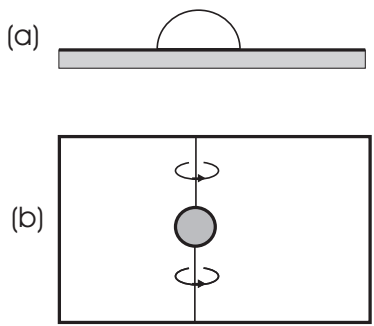}
\caption{Forms of remanent vortex.}
\label{fig6}
\end{center}
\end{figure}

We recall from Section I that the generation of turbulence in $^4$He must rely on the growth of remanent vortices. If the helium has been allowed to settle before the velocity is applied, each remanent vortex must be in metastable equilibrium. It is sometimes suggested that the remanent vortices have the form shown in Fig.~\ref{fig6}(a). However, such a form cannot be in metastable equilibrium in the absence of an imposed flow;  the loop simply collapses into the wall, even if the ends of the loop are pinned at the wall. For the case of a sphere levitated in a box, such as was studied by Schoepe, an allowed remanent vortex is shown in Fig.~\ref{fig6}(b);  the ends of the vortex need not be pinned to protuberances on the sphere or the containing walls.  We have used this form of remanent vortex in the simulations described in the next Section. 

There have been recent reports that the effective mass of oscillating wires and grids in superfluid $^4$He is anomalously large \cite{yano1,charalambous1}, and it has been suggested that this effect might be due to very large density of small remanent vortex loops attached to the surface of the structure (although the effect seems not to be very reproducible). According to our present knowledge, it is hard to understand why such a large density of small remanent vortices can persist, although it is easy to see that a small density of long lengths of remanent vortex might well persist for long periods at very low temperatures, where mutual friction has become very small. For the present we shall assume that the enhanced effective mass has some other cause, and we shall continue to assume that the remanent vortices relevant to the simulations we have performed are few in number. If this assumption proves to be unfounded, our simulations will require modification.

\subsection{Comments on the nature of quasi-classical behaviour.}

We have mentioned that one aim of this work is to discover the extent to which turbulent flow of a superfluid past an obstacle mimics the corresponding classical flow. Before we proceed further we must examine the extent to which this can really be the case.  

One of the clearest and most widely discussed examples of this tendency for turbulent superflow to mimic its classical counterpart is provided by steady flow through a grid \cite{stalp1,vinen3}. Here the relevant experiments were carried out with $^4$He above 1K, where there is at least a small fraction of normal fluid.  A common description of the classical version of this flow at high Reynolds number supposes that flow through the grid leads to the injection of energy into eddies with size of order the mesh of the grid, followed by transfer of this energy in a Richardson inertial cascade to smaller and smaller eddies, until  it can be dissipated by viscosity. A similar picture is given for the quantum case, except that viscous dissipation is replaced by dissipation due to a combination of normal-fluid viscosity and mutual friction, this dissipation taking place on a length scale comparable with the spacing between vortex lines. It is widely believed \cite{vinen1} that this picture continues to hold at very low temperatures, for both $^4$He and $^3$He-B, except that the dissipation is now due to emission of excitations (phonons or, in the case of $^3$He-B,  Caroli-Matricon quasi-particles in the vortex cores). This emission is associated with either vortex reconnections or vortex (Kelvin) waves of high frequency, the scale on which dissipation is occurring being then less than the average vortex separation.    

It is important to recognize that there must be an important difference between the classical and quantum cases, at least in principle.  In the classical case there can be large eddies even in the absence of small eddies, so that we can imagine that large eddies are produced in the neighbourhood of the grid,  and that the smaller eddies are produced only later in the flow by decay of the large eddies.  This cannot be the case with a superfluid, since the large eddies can then exist only as a result of the partial polarization of a tangle of vortex lines; in other words the large eddies can exist only in the presence of very small eddies associated with the vortex tangle. Thus flow through the grid must produce a tangle of vortex lines on a scale less than the mesh of the grid before it can produce eddies on a scale equal to the mesh of the grid. 

For grids formed from holes drilled in a sheet the nature of the classical flow, shown in Fig.~\ref{fig4}, appears to be such that both large and small eddies are produced simultaneously, at least at large enough values of the Reynolds number: the jets emerging from each hole involve vortex sheets, which imply motion on a large range of length scales. The quantum analogue of a vortex sheet is a row of quantum vortices, so that in the quantum case production of turbulence on the scale of the quantized vortex lines may occur in parallel with the production of large scale motion. But in the case of classical flow past a cylinder, or flow through a grid formed from such cylinders, the transition to turbulence seems to involve only the production of large eddies, without the simultaneous production of small eddies, so that quantum analogue must be different. In any case production by a grid of a wide range of eddy sizes might lead to a situation where energy is being injected at a significant rate over the whole of what ought to be the inertial range of wave numbers,  with consequent  departures from the Kolmogorov spectrum.  This appears not to be the case in either the classical or the quantum cases,  so that the injected energy associated with such a wide range of wave numbers must be relatively small.  

\subsection{Simulations with a sphere of radius 100 $\mu$m.}

\begin{figure}[ht]
\begin{center}
\includegraphics[width=0.35\textwidth]{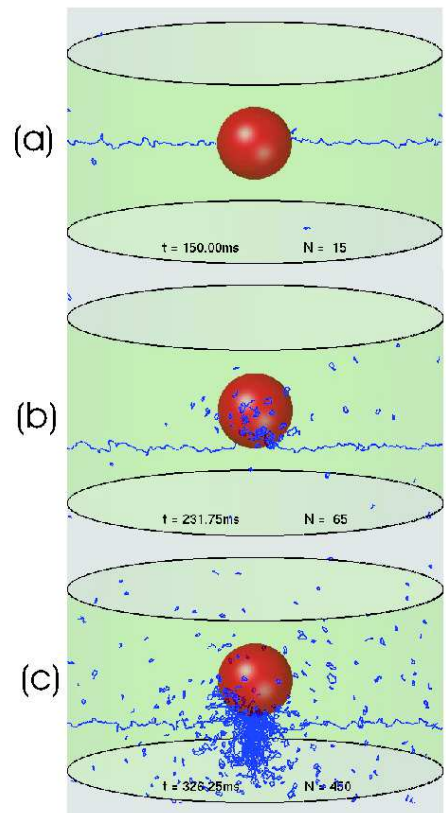}
\caption{A simulation of the development of turbulence around a sphere of radius 100 $\mu$m exposed to an oscillating superflow ($U=150$ mm s$^{-1}$) at a frequency of 200 Hz in $^4$He. }
\label{fig7}
\end{center}
\end{figure}

So far we have had the opportunity to carry out only limited simulations, but the results are nevertheless interesting and point the way towards more extensive studies that we hope to perform in the future. The simulations that we report relate for the most part to the situation depicted in Fig.~\ref{fig6}(b), with a sinusoidal oscillating flow, velocity amplitude $U$, in a direction perpendicular to the unperturbed remanent vortices. They are based on a full Biot-Savart treatment of the vortex filament model, with reconnections assumed to take place when two elements of vortex approach each other within a distance equal to the mesh size in the simulation, but with the proviso that the reconnection must result in a reduction in the total length of line.  

Typical results for a smooth sphere of radius 100 $\mu$m are shown in figure \ref{fig7}; this size of sphere is similar to that used in the experiments of Schoepe \textit{et al}, although their sphere was rather rough. Soon after the oscillation is established (a), Kelvin waves appear on the remanent vortex,  with a dominant wavelength of roughly 40 $\mu$m, which corresponds to the frequency of 200 Hz. Continued oscillation leads to increasing Kelvin-wave amplitudes, with increased non-linear coupling to other wave numbers. Then at sufficiently large Kelvin-wave amplitudes reconnections occur, resulting in the appearance of free vortex rings and vortex loops attached to the sphere (b). In due course (c), a region of strong turbulence appears on one or other side of the sphere (in the direction of the flow), the turbulence resulting apparently from loops being pulled out from the surface of the sphere.  

\begin{figure}[ht]
\begin{center}
\includegraphics[width=0.45\textwidth]{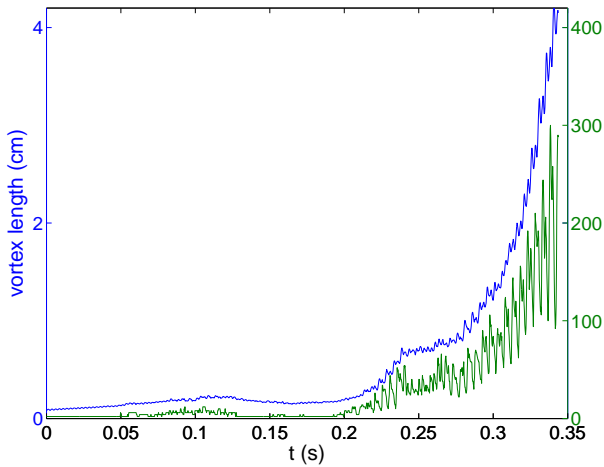}
\caption{The development with time of the total length of vortex line (upper, blue, line; left hand vertical axis) and the number of vortex attachments to the sphere (lower, green, line; right hand vertical axis) for the simulation shown in Fig.~\ref{fig7}.}
\label{fig8}
\end{center}
\end{figure}

This region of strong turbulence is reminiscent of a classical turbulent wake. However, there is a serious reservation. The vortex lines in the wake appear to be completely random, so that, in contrast to the classical case, they generate no large scale motion that can cancel out the potential flow to the rear of the sphere. 

At this point we must refer to Fig.~\ref{fig8}, which shows the development with time of both the total length of vortex line in the region near the sphere and of the number of attached vortices. We see immediately that over the time during which the simulations were carried out the system had not reached a steady state; both the total length of vortex line near the sphere and of the number of attached vortices were still growing with time. Unfortunately, available computer time did not allow any  continuation of the simulation to much larger times.  Therefore we can draw only limited conclusions from the simulations, as follows.  

 Let us first consider the magnitude of the drag on the sphere due to the wake shown in Fig.~\ref{fig7}. This drag is due to the pressure distribution at the surface of the sphere, and we shall regard this pressure as arising from two contributions: the first, $p_1$, due to flow on a scale larger than the vortex spacing; and the second, $p_2$, due to flow on a scale less than the vortex spacing. The fact that there is no cancellation of the large scale potential flow to the rear of the sphere means that the total contribution to the drag from $p_1$ is zero, leaving only the contribution from $p_2$. The major contribution to $p_2$ must come from the high flow velocities near the cores of the vortex lines that are attached to the sphere, and it can therefore be regarded as arising from the tension in these attached lines (the tension being equal to $(\rho_s\kappa^2/4\pi)\ln(\ell/\xi)$), where $\ell$ is the vortex line spacing, and $\xi$ is the radius of the vortex core).  

It is easily shown that the maximum number of attached vortices shown in Fig.~\ref{fig8} leads to a drag coefficient that is of order $8\times 10^{-3}$, which is much less than the observed value of order unity.  Thus the wake does not have a form similar to that in a classical fluid, partly because, as we have already noted, it does not incorporate any appropriate large-scale flow, and partly because, as we now see, it fails to give the correct drag coefficient. There are two possible reasons.  First, the wake has not had sufficient time to develop. And secondly the simulations are in some way unrealistic. If the first reason is relevant then there may be a connection with our comment towards the end of Section IVC that the production of a random tangle of vortex line might need to precede the establishment of a large-scale quasi-classical flow pattern.  The second reason might be connected with the smooth character of the sphere used in the simulations. Perhaps it needs to be rough in order to have quasi-classical behaviour (the sphere used in the experiments by Schoepe \textit{et al} was indeed very rough). Further progress must await the completion of further simulations. The way in which these simulations should proceed is, however, now much clearer. 

It is interesting to ask whether the steady-state drag on a sphere in the quantum case could arise entirely from the contribution $p_2$ to the pressure over the surface of the sphere. In principle, it could, but there seems no reason why this mechanism should lead to a drag coefficient of order unity. In a later section we shall argue that in the case of an oscillating grid in $^3$He-B the dominant contribution to the drag does indeed come from $p_2$, but that the drag coefficient does not then have the classical value. We emphasize our belief that, because in $^4$He the observed drag coefficient is close to unity, most of the drag in that case arises from $p_1$. Confirmation that this belief is correct must come from simulations that extend to times large enough to exhibit a steady state.   

Although in our simulations a wake can sometimes appear in successive half cycles on opposite sides of the sphere, as in the classical case, there seems to be a tendency for it to form more strongly on one side, although the favoured side seems to be random. Further study is required to establish whether this is a real effect, or whether, for example, it is linked to the simulated behaviour not having reached a steady state. The appearance of the wake does not depend on the particular form of nucleating vortex; other forms of nucleating vortex lead to the same type of wake at similar velocities. We note also that the wake extends only a small distance from the sphere; there is no indication of its being cast off from the sphere periodically. In the absence of turbulent regions being cast off from the sphere, dissipation inherent in the drag must occur through two processes: the occasional loss of a vortex ring from the turbulent region; and decay of the turbulence into phonons. In reality, this latter decay process must involve the flow of energy into smaller and smaller length scales and ultimate dissipation into phonons \cite{vinen1}; in the simulations it is associated with the flow of energy into vortex structures that are smaller than the spatial resolution of the simulations. 

The wake does not form if the velocity falls below approximately 120 mm s$^{-1}$, although some vortex rings are still produced by reconnections. Furthermore,  the formation of the wake is not hysteretic: if the wake is established at a velocity exceeding 120 mm s$^{-1}$, and the velocity is reduced below 120 mm s$^{-1}$,  the wake disappears. The velocity of 120 mm s$^{-1}$ is a factor of about three greater than the critical velocity observed by Schoepe's group (Fig.~\ref{fig1}). The production of some vortex rings at the lower velocities implies some drag, but estimates of the energy loss associated with these rings show that this drag is probably too small to be observed. 

As we have noted, our computational studies are still at an early stage of development, and there is the possibility that the results obtained so far are misleading. We emphasize particularly that the development of the turbulent wake shown in Fig.~\ref{fig7} has not been followed to times sufficiently large to ensure that a steady state has been achieved, and that no account has been taken of any roughness of the surface of the sphere. We shall need to learn how to incorporate this roughness into the simulations.  

\subsection{Simulations with a smaller sphere;  other structures}

As we saw in Section II, some of the experiments relate to structures (wires and grids) that are significantly smaller in relevant scale than the 100 $\mu$m sphere. We have therefore carried out simulations with a smaller (smooth) sphere, radius 10 $\mu$m, at a velocity of 150 mm s$^{-1}$ and frequency of 1 kHz, The higher frequency is chosen to match that relevant to these smaller structures.  

We found that there is some tendency to produce a turbulent wake (this time, actually, on both sides of the sphere), but that the wake was much less clearly defined (again we must express a reservation about this simulation because it did not proceed to a steady state). This result is not surprising because quasi-classical behaviour requires the presence of many quantized vortices on a scale of the classical eddies.  For very small structures this condition may not be satisfied.  At first sight one might then expect the drag to be rather different in form from that observed with larger structures.  The smallest structure investigated in experiments so far in $^4$He has been a wire of diameter 2.5 $\mu$m \cite{yano1}. Surprisingly, the experimentally observed drag still appears to be classical in form, although the drag coefficient  is only about 0.17,  which can be compared with the minimum classical value for a cylinder of about 0.3 (Fig.~\ref{fig3}).  

Extensive experiments have been reported on the behaviour of vibrating grids in $^4$He; the grids are formed by making a regular pattern of square holes ($\sim$ 100$\mu$m $\times$ 100$\mu$m) in a thin sheet, the width of the relatively thin opaque strips separating the holes being 20.8 $\mu$m. The drag coefficient, evaluated with the area $A$ equal to that of the opaque part,  proves to be about 0.29. For a thin strip,  which might be expected to behave like a disc at right angles to the flow, we would expect that $C_D$ would be rather close to  unity\cite{prandtl1}. Thus again we find that the drag coefficient associated with a small structure is somewhat less than the classical value.

\subsection{Critical velocities}

We have already noted that the critical velocity at which the strongly turbulent wake appears in the simulations for the 100 $\mu$m sphere is larger than that observed by Schoepe \textit{et al} by a factor of about three. At present we do not know the reason, although we wonder whether it is due to the roughness of the Schoepe sphere.  An interesting feature of the simulations is that the critical velocity seems to be largely independent of the form of the nucleating remanent vortex. If this feature is always found, it could account for the fact that, as we noted in Section II, observed critical velocities are so reproducible from one experiment to another. A challenge for the future is to understand the physics underlying the critical velocity, which is not obvious from the simulations.  

\section{Oscillating structures in $^3$He-B at very low temperatures}

Two structures have been studied in this case:  a vibrating wire, of typical diameter 13$\mu$m \cite{fisher1}; and a grid, similar in design to that used in the experiments we have described in $^4$He,  but with characteristic dimensions reduced by a factor of about two \cite{bradley2,bradley3}. In the case of the wire, the process of vortex creation seems to occur at  much the same velocity as the Landau critical value for pair breaking (about 8 mm s$^{-1}$);  clearly this complicates the interpretation,  and we shall therefore focus our attention on the grid,  where pair breaking seems not to be playing a significant role.  In the case of the grid, as we have already mentioned,  the critical velocity for vortex creation is not well-defined but seems to be not more than about 1 mm s$^{-1}$, which is smaller than that observed in $^4$He by a factor of at least 20 (Fig.~\ref{fig1}). On the other hand drag coefficients for the grid at high velocities are much larger than in $^4$He,  by a factor again of at least 20 (Table~I).  

We have performed simulations similar to those illustrated in Fig.~\ref{fig7}, but with vortex parameters (quantum of circulation and core size) relevant to $^3$He-B. The results are not noticably affected by this change. This suggests that the mechanism for production of vortex line is quite different in the two cases. We know from the experiments of Parts \textit{et al} \cite{parts1} that the critical velocity for intrinsic vortex creation at a solid boundary  in $^3$He-B can be quite small, and that it falls with increasing roughness of the boundary. The surface of the grid is not well characterized, but it is likely to be very rough in comparison with the surfaces studied by Parts \textit{et al}. Furthermore, the grid has sharp edges at each aperture, which will enhance the local superfluid velocity. It seems reasonable to conclude that the mechanism for vortex production by the oscillating grid in $^3$He-B was intrinsic. Presumably this process involves the production of vortex loops at the surface of the grid,  which then expand under the influence of the superflow relative to the grid.  The rate at which this process produces vortex line is not known, but it could well be much greater than that for extrinsic nucleation in $^4$He. 

We shall try to understand how the drag coefficient can be so large.  We recall first that this drag is associated with the distribution of pressure over the structure, and that this pressure can be regarded as having two contributions: $p_1$, due to flow on a scale larger than the vortex spacing;  and  $p_2$, due to flow on a scale less than the vortex spacing. We suggested that  in the case of $^4$He the contribution $p_1$ is quasi-classical, while the contribution $p_2$ can be neglected. We now suggest very tentatively that in $^3$He-B the contribution $p_2$ is dominant and accounts for the large drag.

We first make the comment that in order for our supposition to be correct that the drag in $^4$He is due primarily to $p_1$ there must be a sufficient density of vortex lines behind the sphere that the polarization of these lines  can lead to a flow that mimics a classical wake. At the same time this minimum density must not be so large that the effect of $p_2$ can become significant. Thus if, as we suggest, $p_2$ is important in the case of $^3$He-B, then this minimum must be greatly exceeded. We suggest that, like the low critical velocity, this situation has its origin in intrinsic nucleation. We suggest in fact that intrinsic nucleation produces vortex line at a rate that greatly exceeds that due to extrinsic nucleation in $^4$He. If the drag on the grid used by the Lancaster group in $^3$He-B were due entirely to the tension in attached vortices, the value of the vortex spacing at a grid velocity of 10 mm s$^{-1}$ would have to be about 2 microns, which is not unreasonable. There may still be a contribution to the drag from the pressure distribution $p_1$,  but it is swamped by that due to $p_2$.

We now comment on the picture of turbulence produced by an oscillating grid in $^3$He-B suggested by the Lancaster group \cite{bradley2,bradley3}. This is that there is a production of vortex rings in the immediate neighbourhood of the grid; these rings fly away from the grid, but, at high enough grid velocities, they reach a density at a point behind the grid such that they interact to give a turbulent field. The production of rings near the grid is seen as arising from the excitation of Kelvin waves on remanent vortices, with subsequent reconnections, as occurs in the simulations that we displayed in Fig.~\ref{fig7}(b). We have argued, however, that the critical velocities observed in $^3$He-B are too small for this process to occur, and that in fact vortex line is created by intrinsic processes. Such processes will presumably lead first to the generation of vortex loops attached to the grid, but these loops may be pulled out into the flow, so that reconnections can lead to the formation of vortex rings. In due course these rings will interact to give a turbulent field. Thus the essential features suggested by the Lancaster group remain valid.  The picture of independent rings must break down at high grid velocities.    

In a recent paper \cite{bradley3} the Lancaster group has reported observations of the decay of turbulence after oscillation of their grid is stopped. They suggest that the decay may be similar to that observed in the wake of a steadily moving grid \cite{stalp1}, with a maximum eddy size of order 2 mm,  which is much larger than the mesh of the grid. This result suggests that the flow produced by the oscillating grid is rather different from that observed in a classical fluid by Voropayev and Fernando \cite{voropayev1}, for which the maximum eddy size seemed to be of order the mesh of the grid. This conclusion is not inconsistent with our view that  the production of the turbulence is not similar to that occurring in a classical fluid, but we cannot go further at this stage. It would be interesting to examine whether there is evidence in the Lancaster experiments for a gradual diffusion of vorticity away from the grid,  similar to that described in Section IIIC, during the period when the grid is driven. We remark that in the Lancaster $^3$He experiments the grid oscillates in an unconfined region, whereas in the classical experiments and in the experiments with $^4$He the grid oscillates within a confining and closely fitting cylinder, the axis of which is perpendicular to the plane of the grid. It seems possible therefore that flow of the $^3$He through the moving grid can be accompanied by significant flow round its outer edge, so that the resulting turbulence may incorporate characteristics similar to those produced by an oscillating  disc with size equal to that of the whole grid. 

\section{Summary and Conclusions}

We have discussed the likely forms of quantum turbulence produced by various oscillating structures at very low temperatures, in the light of what is known in analogous classical cases. We have paid particular attention to observed values of drag coefficients, and to the results of simulations of the quantum cases. Much remains to be studied and understood, but tentatively we conclude that oscillating structures in superfluid $^4$He behave in a way that is reminiscent of classical behaviour, with some modification when the size of the structure is not large in comparison with the characteristic length scale (the quantized vortex spacing) in the quantum turbulence. In the case of $^3$He-B  the behaviour seems to be quite different from the classical cases, and we suggest that this is associated with very fast intrinsic nucleation of the quantized vortices.

As we have emphasized several times,  the simulations that we have carried out so far are quite incomplete.  However, the completion of more satisfactory simulations will take time, especially as some will require the development of new techniques to take account of surface roughness. Therefore we think that it is right to publish our preliminary conclusions in this paper.

\begin{acknowledgements}
We are grateful to Professor Russell Donnelly for helpful discussions on the behaviour of oscillating structures in classical fluids and for showing us his own experimental results on an oscillating sphere in water in advance of publication. We are grateful also to Shaun Fisher for his comments on an early version of this paper. MT acknowledges support from a Grant-in-Aid for Scientific Research from JSPS (Grant No. 18340109) and a Grant-in-Aid for Scientific Research on Priority Areas from MEXT (Grant No. 17071008). RH acknowledges the receipt of a grant from the Foundation for the Advancement of Technology in Finland.  
\end{acknowledgements}

\section*{APPENDIX}

We discuss how to identify critical velocities for transition to turbulent flow from experimental plots of  velocity against drag force.  

\begin{figure}[ht]
\begin{center}
\includegraphics[width=0.4\textwidth]{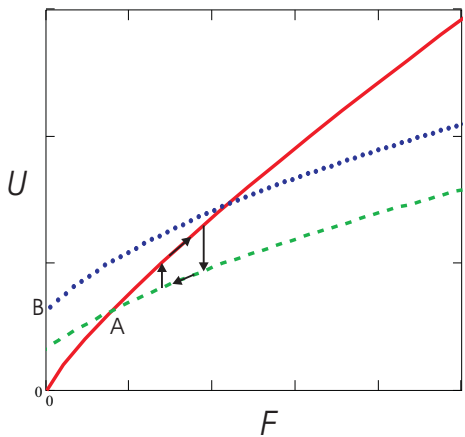}
\caption{Schematic plots of velocity against drag force. }
\label{fig9}
\end{center}
\end{figure}

We refer to Fig.~\ref{fig9}, in which we plot schematically various dependences of velocity on drag force. We recall that in a typical experiment the velocity of the structure is measured over a range of particular applied forces. The solid (red) line shows how the drag might vary for the case of potential flow of the superfluid component. In this case the drag is due to losses within the oscillating structure itself (due, for example, to internal friction) and, at a small finite temperature, to the residual normal fluid. The dotted (blue) line shows how the component of the drag due to turbulence in the superfluid component might vary with velocity. Let us now assume that the drag represented by the solid (red) line is not changed by the onset of turbulence in the superfluid component. In that case the total drag in the presence of a turbulent superfluid component at a particular velocity is obtained by adding together the two drags represented by the solid (red) and dotted (blue) curves. The result is the broken (green) line (note, however, that the part of this line to the left of the point A is unrealistic because it relates to an extrapolation of the dotted (blue) line to negative values of $F$). Suppose now that the applied force is gradually increased from zero. At low values of the velocity the flow of the superfluid component remains potential, and the response of the system follows the solid (red) line. At high velocities flow of the superfluid component is turbulent,  and the response follows the broken (green) line. The transition between one regime and the other takes place in the neighbourhood of the point of intersection A, and the detailed behaviour in this region is not obvious. One possibility that seems sometimes to occur in practice is that the transition takes place in a hysteretic manner, as shown by the arrowed {black) lines, and in this case there is sometimes switching between the turbulent and potential flow regimes. In other cases there seems to be a smooth transition from the solid (red) line to the broken (green) line. In any case, we can identify the point of intersection A by suitable extrapolation. We see immediately that the velocity corresponding to the point A is equal to the velocity B at which the dotted (blue) line intersects the velocity axis. It follows that this velocity is equal to the minimum critical velocity for transition to turbulence in the superfluid component at zero temperature (in the absence of any dissipation not associated with this turbulence). We emphasize, however, that the critical velocity may be larger than this minimum value, as would be the case if there were hysteresis of the type shown by the arrowed (black) lines. Furthermore, our conclusion holds only if the damping associated with the solid (red) line is not changed by the transition to turbulence in the superfluid (the importance of this condition was understood by J\"ager \textit{et al} \cite{schoepe5}). It seems very likely that this is indeed the case if this damping is due to losses within the oscillating structure itself, but it may not be accurately the case for a contribution to the damping from the normal fluid, in which the flow pattern (or the paths of the quasi-particles in a ballistic regime) might well be modified by the mutual friction associated with the superfluid turbulence.

\end{document}